\documentclass{cupconf}

 \checkfont{eurm10}
 \iffontfound
   \IfFileExists{upmath.sty}
     {\typeout{^^JFound AMS Euler Roman fonts on the system,
                  using the 'upmath' package.^^J}%
      \usepackage{upmath}}
     {\typeout{^^JFound AMS Euler Roman fonts on the system, but you
                  dont seem to have the}%
      \typeout{'upmath' package installed. cupconf.cls can take advantage
                of these fonts,^^Jif you use 'upmath' package.^^J}%
     }
 \else
 \fi

 \checkfont{msam10}
 \iffontfound
   \IfFileExists{amssymb.sty}
     {\typeout{^^JFound AMS Symbol fonts on the system, using the
               'amssymb' package.^^J}%
      \usepackage{amssymb}%
      \let\le=\leqslant  
      \let\ge=\geqslant  
     }{}
 \fi

 \IfFileExists{amsbsy.sty}
   {\typeout{^^JFound the 'amsbsy' package on the system, using it.^^J}%
    \usepackage{amsbsy}}
   {}

\newsavebox{\astrutbox}
\sbox{\astrutbox}{\rule[-5pt]{0pt}{20pt}}

\newcommand{\Msun}{~M_\odot}
\newcommand{\msun}{M_\odot}

\newcommand{\kms}{\rm ~km~s^{-1}}

\newcommand{\ml}{~\Msun ~\rm yr^{-1}}

\newcommand{\lsim}{\!\!\!\phantom{\le}\smash{\buildrel{}\over
 {\lower2.5dd\hbox{$\buildrel{\lower2dd\hbox{$\displaystyle<$}}\over
                              \sim$}}}\,\,}

\newcommand{\gsim}{\!\!\!\phantom{\ge}\smash{\buildrel{}\over
 {\lower2.5dd\hbox{$\buildrel{\lower2dd\hbox{$\displaystyle>$}}\over
                              \sim$}}}\,\,}
\usepackage{graphicx}

\title[From progenitor to afterlife]{From progenitor to afterlife}

\author[R. A. Chevalier]{Roger A. Chevalier}

\affiliation{Department of Astronomy, University of Virginia,
P.O. Box 400325, Charlottesville, VA 22904, USA\\}

\begin{document}

\maketitle

\begin{abstract}
The sequence of massive star supernova
types IIP (plateau light curve), IIL (linear light curve), IIb,
IIn (narrow line), Ib, and Ic roughly represents a sequence of
increasing mass loss during the stellar evolution.
The mass loss affects the velocity distribution of the ejecta
composition; in particular, only the IIP's typically end up
with H moving at low velocity.
Radio and X-ray observations of extragalactic supernovae show
varying mass loss properties that are
in line with expectations for the progenitor stars.
For young supernova remnants, pulsar wind nebulae 
and circumstellar interaction provide 
probes of the inner ejecta and higher velocity ejecta, respectively.
Among the young remnants,
there is evidence for supernovae over a range of types, including
those that exploded with much of the H envelope present
(Crab Nebula, 3C 58, 0540--69) and those that exploded after
having lost most of their H envelope (Cas A, G292.0+1.8).

\end{abstract}

\firstsection 
\section{Introduction: Core Collapse Supernovae}

Core collapse supernovae show considerable diversity among their
properties.
A basic observational division is into the SNe II 
(Type II supernovae), which have
hydrogen in their spectra, and SNe Ib/c, which do not (or have weak
hydrogen lines).
The reason for the difference is that the progenitors of the SN Ib/c
have lost their H envelopes, and perhaps more, during their evolution
leading up to the supernova.
The mass loss can occur either through the winds from a single star
or can be aided by interaction with a binary companion.

The SNe II show strong diversity themselves.
Their observational classification is based on a variety of factors,
but it is clear that presupernova mass
loss plays a significant role in determining the type.
Two types are distinguished by their light curves: IIP (plateau)
and IIL (linear).
Models of Type IIP light curves have long showed that the likely
progenitors of the SNe IIP are the red supergiants that end their
lives with most of their H envelopes retained 
(\cite[Grasberg et al. 1971]{GIN71}, \cite[Chevalier 1976]{Chev76}).
The plateau phase of the light curve is due to the internal energy
deposited by the initial explosion.
This progenitor hypothesis has been directly confirmed by
observations of the progenitors of a number of SNe IIP 
(\cite[Hendry et al. 2006]{Hen06} and references therein).
While the SNe IIP might explode with a hydrogen envelope of 
$\sim10\Msun$, the more rapid decline of the SNe IIL imply that
they explode with 
an envelope of $\sim1\Msun$ (\cite[Blinnikov \& Bartunov 1993]{BB93}).
Because of higher rates of mass loss for more luminous stars, the
reduced H envelope is expected to occur for single stars with
initial masses of $\gsim20\Msun$.
Alternatively, mass loss in a binary system could play a role in the
reduced envelope mass.

The prototype of the SNe IIb was SN 1993J, which made a transition
from a Type II at early times to a Type Ib/c at late times,
based on spectroscopic observations.
The H envelope mass required for SN 1993J was $\sim0.2\Msun$
(\cite[Woosley et al. 1994]{Woo94}).
For this to occur in a single star requires special timing, so a binary
origin is preferred.
A likely binary companion for SN 1993J has been directly
observed (\cite[Maund et al. 2005]{Mau05}).

SNe IIn have the spectroscopic feature of narrow emission lines 
(\cite[Schlegel 1990]{Sch90}),
typically H$\alpha$, which indicates that
circumstellar interaction plays a role in the emission from
early times.
A supernova can be a Type IIn and another type; e.g., SN 1998S was
both a IIn and IIL.
Because of the strong circumstellar interaction, it can be difficult
to determine the nature of the photospheric emission in a SN IIn.
The H emission from circumstellar interaction implies strong
mass loss before the supernova in a SN IIn, so the H envelope is
likely to be depleted at the time of the supernova.

The most noteworthy peculiar SN II is the nearby SN 1987A, which was
relatively compact at the time of the explosion although it had a 
massive H envelope.
The best explanation for the explosion as a blue supergiant star and
the axisymmetric ring features around it is probably that it was
in a binary system (\cite[Podsiadlowski 1992]{Pod92}).

The Type Ib/c supernovae are believed to be H-poor Wolf-Rayet stars
at the time of their explosion.
There is some 
observational evidence that SNe Ib, which have  He lines, can have a
small amount of high velocity H at the time of the explosion 
(\cite[Elmhamdi et al. 2006]{Elm06}).
Although the presence of a H envelope in a massive star typically
leads to the formation of a red supergiant in the late evolutionary
stages, a small amount of H mass ($\lsim0.01\Msun$) is not expected to
support an extended envelope.

These considerations show that a major factor in the determination of
supernova type is the amount of H left in the envelope at the time of
the supernova.
If the H envelope mass is greater than the core mass, then the core
is effectively decelerated by the envelope and there is mixing between
them by Rayleigh-Taylor instabilities.  
There is not only the outward mixing of heavy elements, but also
the inward mixing of H to low velocities.
This can be directly observed in the late spectra of SNe IIP;
e.g., late spectra of SN 1999em showed H moving at several $100\kms$ 
(\cite[Elmhamdi et al. 2003]{Elm03}).

The relative numbers of the different kinds of supernovae is uncertain
from an observational point of view.
If all the supernovae came from single stars, the stellar mass function
was the Salpeter function ($n(M)\propto M^{-2.35}$), and Type IIP supernovae
came from $8-20\Msun$ stars, Type IIL $20-25\Msun$, and Type Ib/c
$>25\Msun$, the relative fractions of IIP:IIL:Ib/c would be 0.71:0.08:0.21.
Binary evolution could increase the relative number of IIL and Ib/c
events (\cite[Nomoto et al. 1996]{Nom96}).
However, SNe IIP are likely to be an important component of the core
collapse supernovae.

After the explosion of the progenitor star, the event has an afterlife
in two ways: through its interaction with the surrounding medium and
through the possible activity of a central compact remnant (neutron
star or black hole).
I will discuss how the expectations for the afterlife phases depend
on the supernova type and what can be learned about these phases from
observations.
The early circumstellar interaction observed in extragalactic
supernovae is discussed in Section 2, the pulsar wind nebula
expansion inside the supernova in Section 3, and circumstellar
interaction in young remnants in Section 4.
More details on the material in Sections 3 and 4 can be found in
\cite{Chev05}.
Section 5 contains a discussion of future prospects.

\section{Early Circumstellar Interaction}\label{sec:csm}

Circumstellar interaction begins soon after the supernova shock wave has
emerged from the progenitor star.
The radiation dominated
shock front accelerates out the outer edge of the star until the point
where radiative losses halt the acceleration process.
Because more compact stars have a larger density contrast between the
average density and the photospheric density, the shock waves attain
higher velocities in more compact stars.
The shock break-out radiation accelerates the gas out ahead of the
shock, so that the shock front in fact disappears.
However, the velocity of the radiatively accelerated gas declines
with radius ($\propto r^{-2}$), so that a viscous shock eventually
forms.
The shocked region is driven by the supernova gas, which has a
steep power law profile in the region where shock acceleration has
occurred.
The interaction region is bounded by a reverse shock, where the
supernova gas is shocked, on the inside and a forward shock, where
the circumstellar gas is shocked, on the outside.
If both the supernova density profile and the surrounding wind density
profile can be described as power laws in radius, the structure
and evolution of the interaction region can be described by a
self-similar solution (\cite[Chevalier 1982]{Che82}).

The early interaction in extragalactic supernovae can be observed in
a number of ways: radio emission from shock accelerated electrons,
X-ray emission from hot gas and nonthermal processes, optical emission
from cooling shock waves and radiatively heated gas, and infrared emission
from radiatively or shock heated dust grains.
Radio is the best marker of interaction because it has been observed from
all the types of massive star supernovae (\cite[Weiler et al. 2005]{Wei05}).
On the other hand, radio emission has never been detected from SNe Ia.

Although there is not a good understanding of particle  acceleration and
magnetic field generation associated with shocked regions, simple models that
assume that some fraction of the postshock energy density goes into
relativistic electrons and magnetic field do a reasonable job of 
reproducing the observed evolution of radio supernovae.
Shock compression of a stellar wind magnetic field is typically not adequate
(unless the magnetic field completely dominates the wind energy flux),
so field amplification in the shocked region is required.
Possible mechanisms for amplification are hydrodynamic instabilities
in the shocked region or field amplification associated with cosmic
ray driven turbulence in the shock wave (\cite[Bell 2004]{Bel04}).
An additional factor
in radio light curves is early low frequency absorption of the radio
emission, as is typically observed.
The expected mechanisms are synchrotron self-absorption and free-free
absorption by unshocked stellar wind gas.

\begin{figure}
\begin{center}
\includegraphics[scale=0.6,origin=c]{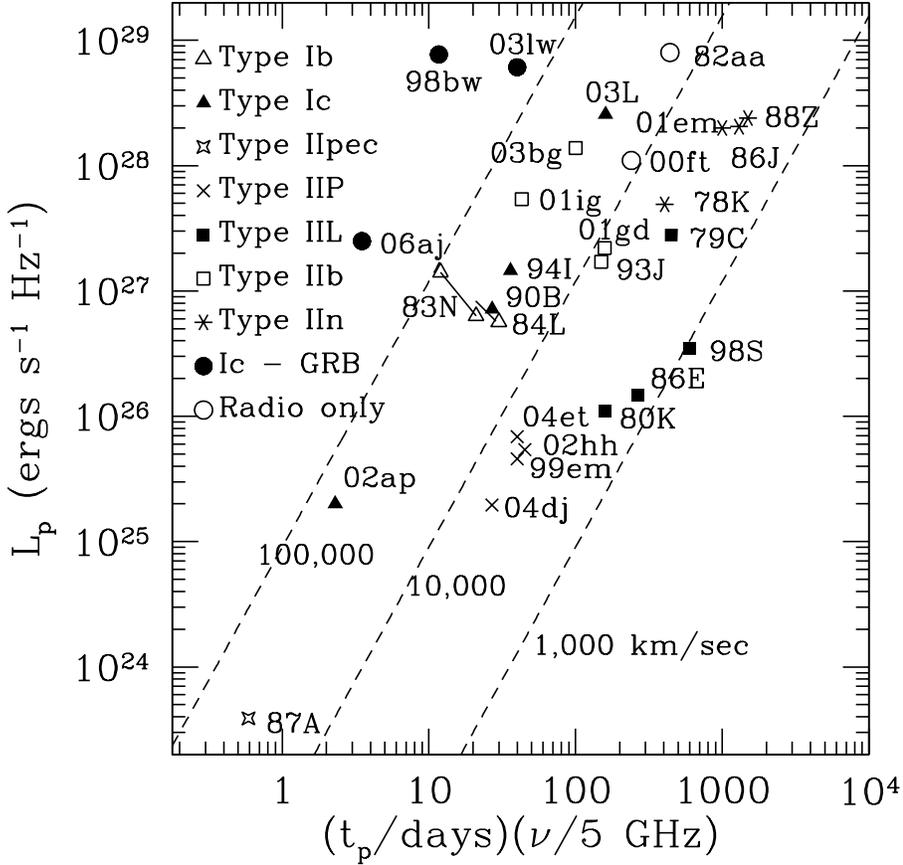}
\end{center}
\caption[]{Peak radio spectral luminosity vs. the product of the time
of peak and the frequency of the measurement.
The observed supernovae are designated by the last two digits of the
year and the letter, and the Types are indicated by the symbols.
The dashed lines show the mean velocity of the radio shell if
synchrotron self-absorption is responsible for the flux peak;
a value of the energy index $p=2.5$ is assumed.
This is an update of Fig. 4 of \cite{Che98}. }
\label{fig1}
\end{figure}

A basic aspect of a radio light curve is thus the peak luminosity and
the time of the peak.
Values are shown in Fig. \ref{fig1} for those supernovae that have light curves.
The dashed lines in Fig. \ref{fig1} result from a synchrotron self-absorption
interpretation of the rise at radio wavelengths; the assumption of
equipartition of energy between relativistic electrons and magnetic
fields gives the radius, and thus the velocity, of the radio emitting
region.
Although equipartition is by no means guaranteed, the results are not
sensitive to this assumption.
If some process other than synchrotron self-absorption is the dominant
absorption process, the radio turn-on is further delayed and the indicated velocity
is lower than actually present in the supernova.
This is the reason that some of the SNe II have very low apparent velocities;
the turn-on is likely due to free-free absorption.

The supernovae divide themselves into 3 main regions with regard to
velocity.
The SNe II have the lowest velocities, although not actually as low as indicated
by Fig. \ref{fig1}.
The SNe Ib/c probably are dominated by 
synchrotron self-absorption and their typical velocities
are $\sim30,000\kms$.
There are 3 reasons for the higher velocities in the SNe Ib/c vs. SNe II:
shock acceleration during the supernova 
continues to higher velocities in the SNe Ib/c because
of the more compact progenitors, the lower circumstellar densities around
the SNe Ib/c give less deceleration of the interaction region, and
the SNe Ib/c typically have lower ejecta masses than, but similar energies
to, the SNe II so the ejecta have higher mean velocities.

Fig. \ref{fig1} also shows the relatively nearby GRB (gamma-ray burst) -- SN Ic 
associations.
They require relativistic or semi-relativistic velocities and thus 
distinguish themselves from the normal SNe Ib/c.
They do not have especially  low ejecta mass 
($\sim6\Msun$ was deduced for SN 1998bw, \cite[Woosley et al. 1999]{WES99}),
so the high velocity must be due either to an extraordinarily large 
supernova energy or to a different source for the emission, a central
GRB engine.

Among the SNe II, the range in radio luminosity is probably due to a
range in mass loss density.
At the low luminosity end are the SNe IIP, which have been detected only
in recent years.
The mass loss rates that are implied by the radio (and X-ray) observations
are consistent with those suggested by stellar evolution calculations,
which are deduced from observations of Galactic stars 
(\cite[Chevalier, Fransson \& Nymark 2006]{CFN06}).
Even over the mass range $10-20\Msun$, there is a considerable range
of mass loss rate.
The existing data are consistent with the expected correlation between
mass loss rate and progenitor mass determined from direct observations,
although there are not yet enough data to provide a good test.

The SNe IIL have a higher mass loss rate, as expected if they
came from single stars.  
If binaries play a role, the expected mass loss rates are not so
clear.
At the high radio luminosity end of the SNe II are the SNe IIn, which
appear to have massive, clumpy circumstellar media.
In order to obtain the high radio, X-ray, and optical luminosities, the
progenitor star must have lost several $\Msun$ within $\sim10^3$
years of the supernova.

Results on mass loss densities are summarized in Table \ref{tab:sn}, where $\dot M$
is the mass loss rate and $v_w$ is the wind velocity of the progenitor;
supernova observations just give the ratio $\dot M/v_w$, so 
the value of $v_w$ has been
assumed.
For the SNe II (except for SN 1987A), 
free-free absorption is the likely absorption process, so
there is a fairly direct estimate of the circumstellar density, although
uncertainties arise because of the dependence of the absorption on the
circumstellar temperature.

For the SNe IIb, Table \ref{tab:sn} just lists the well-studied SN 1993J, which
had a red supergiant progenitor.
Fig. \ref{fig1} shows that SN 2001gd was probably similar.
However, the radio observations of SN 2001ig and SN 2003bg indicate
higher velocity expansion and they probably had Wolf-Rayet star
progenitors (\cite[Ryder et al. 2004]{Ryd04}, \cite[Soderberg et al. 2006]
{Sod06}).
There is presumably a continuous distribution between the SNe IIb, in
which the H lines are clearly visible in spectra, and the SNe Ib, in which 
the H$\alpha$ line is weak.

The position of SN 1987A in Fig. \ref{fig1} is determined by the low luminosity radio
emission that was observed over the first 200 days after the
explosion (\cite[Turtle et al. 1987]{Tur87}).
The initial rise of the radio emission is likely due to 
synchrotron self-absorption, so 
estimates of mass loss density are uncertain, but, if the efficiency
of synchrotron production is similar to that for the SNe Ib/c (see
below), the density is remarkably low (Table \ref{tab:sn}).
This low density is supported by the rapid expansion that
the supernova shock wave made to the time that the first radio imaging
observations were carried out (\cite[Gaensler et al. 1997]{Gae97}).

\begin{table}
  \begin{center}
  \begin{tabular}{ccc}
      SN   & $\dot M$   &   Assumed $v_w$ \\
             & ($\msun$ yr$^{-1}$) &   (km s$^{-1}$)    \\[3pt]
       IIP   & $10^{-6}-10^{-5}$ &   10 \\
       IIL   & $10^{-5}-10^{-4}$ &   10 \\
      93J (IIb)   & $3\times 10^{-5}$ &   10 \\
       IIn   & $\lsim10^{-3}$ &   10 \\
      87A (IIpec)   & $4\times 10^{-8}$ &   500 \\
       Ib/c   & $10^{-6}-10^{-4}$ &   1000 \\
  \end{tabular}
  \caption{Estimates of $\dot M$ for the supernova progenitors}
  \label{tab:sn}
  \end{center}
\end{table}

Since 1990, the radio flux from SN 1987A has been rising because of its
interaction with mass lost during a previous red supergiant stage.
This increase was anticipated by the observation of dense gas that
had been radiatively illuminated and the ensuing interaction has been
observed over a broad wavelength range (\cite[McCray 2005]{McC05}).
The transition from red supergiant to blue supergiant 
explosion 
took $\sim10^4$ years.
The radio light curve of SN 1987A is unusual because of its previous 
red supergiant phase, although few radio supernovae are followed past
an age of 3 years.
Another object that seems to have made a transition to dense gas interaction
is SN 2001em, which was initially observed as a SN Ib/c and within 3 years
made a transition to SN IIn, at which time it was a luminous radio and
X-ray source (\cite[Chugai \& Chevalier 2006]{CC06} and references therein).
In this case, there was apparently a phase of dense mass loss within
$\sim10^3$ years of the supernova explosion which ended before the
supernova occurred.

Like the SNe II, the SNe Ib/c also have a considerable range in peak
radio luminosity (Fig. \ref{fig1}).
The observed range in luminosity is roughly consistent with the observed range
of mass loss densities for Galactic Wolf-Rayet stars if the efficiency
factors (fractions of postshock energy density in magnetic fields and
relativistic electrons) do not vary greatly between objects and are
$\sim0.1$.
In this picture, SN 2002ap and SN 2003L roughly represent the low and
high extremes for the radio luminosities expected for SNe Ib/c.
The low density around SN 2002ap is a factor in the high velocity of
the radio region that is inferred for this source.

In addition to radio emission, X-ray emission has been detected from
essentially all types of massive star supernovae, except perhaps SNe Ib.
However, there is typically only a small amount of data for any
particular supernova, so that the X-ray data can provide a consistency
check on deductions from the radio emission, but do not yield much
additional information on the mass loss properties of the progenitors.
In the case of SNe II, the X-ray emission is likely to be thermal
emission from the shocked ejecta gas.
The interpretation of the emission generally depends on the density
structure of the supernova.
In the case of SNe Ib/c, the thermal interpretation generally does
not produce sufficient luminosity, so nonthermal mechanisms are
indicated (\cite[Chevalier \& Fransson 2006]{CF06}).
Near maximum optical light, inverse Compton emission can be important,
but it cannot explain later emission.
\cite{CF06} suggested that the late emission can be explained by synchrotron
radiation in a scenario where the forward shock wave is cosmic ray
dominated so that the electron energy spectrum flattens at high
energy.
More detailed observations are needed to check on this hypothesis.

Optical emission from circumstellar interaction occurs if the interaction
is sufficiently dense to produce a radiatively cooling reverse shock wave.
It is in this case that a significant fraction of the interaction power
can appear at optical wavelengths.
Thus, optical emission from interaction
is detected from IIn, IIL, and IIb supernovae,
but not from IIP or Ib/c supernovae.
The H$\alpha$ line profiles observed for IIL and IIb supernovae typically
have the boxy shape that is expected for emission from a fairly narrow
region near the reverse shock front.
The SNe IIn have narrow centrally peaked H$\alpha$ emission that is
likely to be from slow shock waves driven into circumstellar clumps,
although a detailed theory for the formation of such lines is not
yet available.

Overall, there is reasonable agreement between the circumstellar media
inferred from supernova observations with what is expected around
the progenitor.
One area of uncertainty is still what are the expectations where binary
interaction has been important for the progenitor.
In addition, the supernova observations are sensitive to clumping in the
circumstellar wind and may provide a method to investigate clumping
in winds from late type stars (e.g., \cite[Weiler et al. 2005]{Wei05}).

\section{Pulsar Wind Nebulae}\label{sec:pwn}

\begin{figure}
\includegraphics[scale=0.95,origin=c]{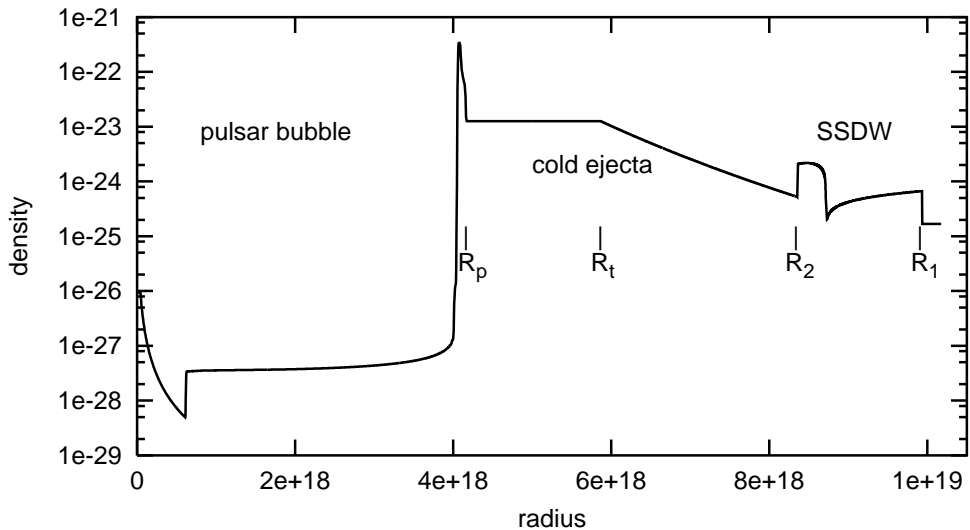}
\caption[]{The density profile  for the interaction of
a pulsar nebula with the host supernova remnant.  The supernova remnant
is modeled as a self-similar driven wave (SSDW) bounded by a forward
shock at $R_1$ and a reverse shock at $R_2$.  The pulsar bubble has
swept up a thin shell of ejecta at $R_p$.
The freely expanding, cold ejecta have an inflection point in the
density at $R_t$.
The reverse shock wave has not yet reached the pulsar bubble.
(from \cite[Blondin, Chevalier \& Frierson 2001]{BCF01})}
\label{fig2}
\end{figure}

Massive stars undergo core collapse at the end of their lives, leading to the 
formation of a neutron star or black hole.
If the neutron star is an active pulsar, the magnetic field and relativistic
particles generated by the pulsar create a bubble within the supernova (Fig. \ref{fig2}).
Because the supernova gas is already freely expanding, the swept up
shell around the bubble accelerates with time and is thus subject to
Rayleigh-Taylor instabilities.
This picture provides a reasonable account of the properties of the
Crab Nebula (\cite[Chevalier 1977]{Chev77}).
A similar model applied to 9 young pulsar nebulae is also in accord with
the observations, provided that the relativistic particles and magnetic
fields are not far from equipartition in the pulsar nebulae 
(\cite[Chevalier 2005]{Chev05}).
In these models for the pulsar nebulae, the initial 
rotation periods of the pulsars
are in the range 10--100 ms.

The fact that the pulsar nebula interacts with the inner part of the
supernova ejecta gives constraints on the supernova type.
As discussed in Section 1, the basic supernova types are related to the
amount of H envelope that is lost leading up to the explosion.
In SNe IIP, with most of their H envelope intact, the core material is
slowed by the envelope material and the reverse shock wave during the 
supernova drives H rich material back towards the center of the supernova.
Once the H envelope has a mass less than that of the core, it does not 
effectively decelerate the core material and ends up at a high velocity.
While the H can give indications of the amount of mass loss, the
heavy element production is related to the initial mass of the star.
Below $\sim12\Msun$, most of the heavy elements synthesized during the
stellar evolution end up in the compact object; above this mass, there
are increasing amount of O (oxygen) and other heavy elements.

When the pulsar bubble expands into the supernova, there is a shock
wave driven in the ejecta.
During the early phases of evolution, the shock is a radiative shock,
which can give optical emission.
The shock emission declines either because the decreasing density and
increasing shock velocity cause a transition to a nonradiative shock
or because the pulsar power declines strongly so that the shell 
expansion tends toward free expansion.
An alternative source of optical emission is photoionization of the
cool gas by the ultraviolet synchrotron radiation from the pulsar bubble.

There are 3 nebulae where the swept up gas has apparently been observed: 
the Crab Nebula, 3C 58, and 0540--69 in the Large Magellanic Cloud.
The Crab is especially well studied and shows filaments that are
primarily composed of H and He, with not much in the way of heavier
elements beyond those present in cosmic abundances.
The small amount of heavy elements suggests a relatively low initial
mass, perhaps $8-10\Msun$ (\cite[Nomoto et al. 1982]{Nom82}).
The average velocity of the optical filaments is $\sim1400\kms$ and the
gas was probably at a lower velocity before being accelerated by the
pulsar bubble.
The low velocity of the H is suggestive of a SN IIP in which the H has
been mixed back toward the center; this type is also consistent with
the estimated initial mass.

The optical filaments in 3C 58 are faint and more difficult to observe,
but show lines of H and N, and velocities up to $\sim1000\kms$
(\cite[Fesen, Kirshner, \& Becker 1988]{FKB88}).
These appear to be swept up ejecta, as opposed to shocked circumstellar
gas, because of their relatively high velocities and the fact that they
appear in projection only over the pulsar nebula.
As in the case of the Crab, the evidence points to a relatively low mass
progenitor and a SN IIP.
Although 3C 58 has typically been identified with SN 1181, there are
a number of lines of evidence that it is actually $\sim2500$ years old 
(\cite[Chevalier 2005]{Chev05}).

The case of 0540--69 is different because strong lines of O and S are present.
The presence of H in the filaments has been controversial and it was
assumed not to be present in \cite{Chev05}, but recent observations definitely
show that it is in the filaments around the pulsar bubble 
(\cite[Serafimovich et al. 2005]{Ser05}),
which have velocities $\sim1000\kms$.
The composition suggests an initial mass $\gsim15\Msun$ and a SN IIP,
which would place the supernova among the higher mass SNe IIP.

\section{Young Remnants}\label{sec:ysnr}

Nearby young remnants with ages up to several 1000 years are expected
to still be interacting with the mass loss region set up by the
progenitor.
However, the mass loss region is probably beyond the region of the
free wind from the progenitor star, so the stellar evolution leading up to
the supernova is important.
When massive stars are on the main sequence, their fast winds can
create large wind bubbles around them.
The lower mass stars have lower mass loss rates, but this is partially
compensated by their longer evolutionary lifetimes.
The wind bubbles eventually slow to $10-20\kms$, which is comparable to
the space velocities of the massive stars, so that they can catch up
to the bubble on one side.

After the main sequence phase, stars enter the red supergiant phase, with
slow ($\sim10\kms$), dense mass loss.
The free wind extends out to the point where the wind ram pressure equals
the pressure in the surrounding medium, $p$, i.e. at
\begin{equation}
r_{RSG}=5.0\left(\dot M\over 5\times 10^{-5}\ml\right)^{1/2}
\left(v_w\over 15\kms\right)^{1/2}
\left(p/k\over 10^4~{\rm cm^{-3}~K}\right)^{-1/2} {\rm~pc},
\label{csrsg}
\end{equation}
where $k$ is Boltzmann's constant.
This shows that the red supergiant wind can extend out $>5$ pc from
the supernova.
The extended wind around SN 1987A was observed as it was illuminated by
the radiation from the supernova, out to a radius $\sim5$ pc
(\cite[Chevalier \& Emmering 1989]{CE89}, \cite[Sugerman et al. 2005]
{Sug05}).

The supernova interaction with a red supergiant wind can last for
1000's of years if the wind is strong and extended, and gives rise
to strong radio and X-ray emission.
The best case of such interaction appears to be the 325 year old remnant
Cas A.
The morphology, expansion rates, and masses are consistent with interaction
with a freely expanding wind (\cite[Chevalier \& Oishi 2003]{CO03},
 \cite[Laming \& Hwang 2003]{LH03}).
The remnant contains slow moving shocked circumstellar clumps, called the
quasi-stationary flocculi, that are H and He rich.
The fact that these give rise to narrow line emission means that Cas A
can be regarded as a very old SN IIn.
Whether it was a SN IIn in its early phases depends on how far back to
the progenitor star the wind extended.
If the progenitor made a transition to a Wolf-Rayet star before
the supernova, it would have initially been a SN Ib/c.
\cite{CO03} 
argued that the wind extended back to near the surface based on 2 points:
there are some fast knots containing H, showing that the progenitor had
some H at the time of the explosion and the formation of very fast cool
knots might be aided by the presence of a dense surrounding wind.
However, the knots with H in Cas A have velocities $\sim10,000\kms$ and
SNe Ib can have H with velocities $\gsim12,000\kms$ 
(\cite[Elmhamdi et al. 2006]{Elm06}).

A general expectation of strong interaction with a red supergiant wind
is that enough of the H envelope has been lost that no H in the ejecta
is expected at low ($\lsim3000\kms$) velocity.
This is the case for the fast ejecta knots in Cas A.
A remnant with strong circumstellar interaction and a pulsar wind
nebula is G292.0+1.8, which also has fast moving knots without H.
Another remnant is 1E 0102.2--7219 in the 
Small Magellanic Cloud, which also has strong interaction
and H-poor fast knots.

As discussed in the previous section, the suggested type of supernova
for the Crab and 3C 58 is a low mass SN IIP.
However, in these cases, there has been no detection of interaction
with the circumstellar medium; quite strong limits have been placed
on X-ray emission from interaction around the Crab nebula 
(\cite[Seward, Gorenstein \& Smith 2006]{SGS06}).
A possible explanation for the low emissivity is that the supernova
shock wave has passed through the red supergiant wind, which did not
extend out far from the progenitor in this case, and is currently
moving in a low density wind bubble left from the main sequence phase.
Some support for this picture comes from the observation of a faint
X-ray shell around the pulsar wind nebula G21.5--0.9 
(\cite[Matheson \& Safi-Harb 2005]{HS05}, 
\cite[Bocchino et al. 2005]{Boc05}).
This remnant was regarded as a pure pulsar nebula, like the Crab,
until long X-ray observations were undertaken with {\it Chandra}.

\cite{Chev05} suggested that the remnant 0540--69  came
from a SN Ib/c based on the apparent rapid expansion of the outer
ejecta, which implied a low circumstellar density in the region
surrounding the progenitor.
However, as discussed in Section 3, H is present in the slow moving
ejecta, which implies a IIP supernova.
In this case, the interaction that is observed at a radius of $6-10$ pc
is probably with the interstellar medium; \cite{Hwa01} estimate that mass of 
X-ray emitting gas is $\sim40\Msun$.
The fact that the X-ray temperature is relatively low for the average
shock velocity suggests that the remnant is interacting with clumps
or clouds.
The problem with the IIP designation is the rapid expansion despite
the expected interaction
with the slow wind from the red supergiant progenitor.
This issue requires more investigation.

\section{Discussion and Conclusions}\label{sec:dc}

There are excellent future prospects for
developing a more complete picture of the massive
star evolution leading up to a supernova and the subsequent
expansion of the supernova into the circumstellar medium.
The increasing number of {\it Hubble Space Telescope} images
of galaxies has improved the prospects for identifying the
progenitor stars of nearby supernovae.
Follow up observations at radio and X-ray wavelengths can
then reveal the mass loss environment for that particular
progenitor.

There is a growing number of young remnants that have
observed
pulsar wind nebulae and/or circumstellar interaction.
Many of these have been well observed at X-ray wavelengths
(owing to {\it Chandra} and {\it XMM}), but are less well
observed at optical and infrared wavelengths.
Infrared observations seem especially important because
a number of the objects have high extinction.
As the amount of information increases, there is the
possibility of looking for correlations between the nature
of the compact object in a remnant and the nature of the
surrounding supernova.
An initial examination of this point (\cite[Chevalier
2005]{Chev05}) did not reveal any correlations.

Along with these endeavors, hydrodynamic modeling of 
the variety of supernova events, along with their interaction
with mass loss, is needed.
The result will be a better understanding of the final
evolution of massive stars and the variety of possible
outcomes.

\begin{acknowledgments}
This research was supported in part by NSF grant
AST-0307366
and NASA grant NAG5-13272.
\end{acknowledgments}


\begin{thebibliography}{}

  \bibitem[Bell (2004)]{Bel04}
    \textsc{Bell, A. R.} 2004
    \textit{MNRAS} \textbf{353}, 550.
    
 \bibitem[Blinnikov \& Bartunov (1993)]{BB93}
    \textsc{Blinnikov, S. I. \& Bartunov, O. S.} 1993
    \textit{A\&A} \textbf{273}, 106.

 \bibitem[Blondin, Chevalier \& Frierson (2001)]{BCF01}
    \textsc{Blondin, J. M., Chevalier, R. A. \& Frierson, D. M.} 2006
    \textit{ApJ} \textbf{563}, 806.
 \bibitem[Bocchino et al. (2005)]{Boc05}
    \textsc{Bocchino, F., van der Swaluw, E., Chevalier, R. \& Bandiera, R.} 2005
    \textit{A\&A} \textbf{442}, 539.

 \bibitem[Chevalier (1976)]{Chev76}
    \textsc{Chevalier, R. A.} 1976
    \textit{ApJ} \textbf{207}, 872.
 \bibitem[Chevalier (1977)]{Chev77}
    \textsc{Chevalier, R. A.} 1977
    In \textit{Supernovae} (ed.\ D. N. Schramm) p.\ 53. Reidel.
 \bibitem[Chevalier (1982)]{Che82}
    \textsc{Chevalier, R. A.} 1982
    \textit{ApJ} \textbf{258}, 790.

 \bibitem[Chevalier (1998)]{Che98}
    \textsc{Chevalier, R. A.} 1998
    \textit{ApJ} \textbf{499}, 810.

 \bibitem[Chevalier (2005)]{Chev05}
    \textsc{Chevalier, R. A.} 2005
    \textit{ApJ} \textbf{619}, 839.

 \bibitem[Chevalier \& Emmering (1989)]{CE98}
    \textsc{Chevalier, R. A. \& Emmering, R. T.} 1989
    \textit{ApJ} \textbf{342}, L75 
 \bibitem[Chevalier \& Fransson (2006)]{CF06}
    \textsc{Chevalier, R. A. \& Fransson, C.} 2006
    \textit{ApJ} \textbf{651}, in press (astro-ph/0607196).
 \bibitem[Chevalier, Fransson \& Nymark (2006)]{CFN06}
    \textsc{Chevalier, R. A., Fransson, C. \& Nymark, T.} 2006
    \textit{ApJ} \textbf{641}, 1029.
 \bibitem[Chevalier \& Oishi (2003)]{CO03}
    \textsc{Chevalier, R. A. \& Oishi, J.} 2003
    \textit{ApJ} \textbf{593}, L23.
 \bibitem[Chugai \& Chevalier (2006)]{CC06}
    \textsc{Chugai, N. N. \& Chevalier, R. A.} 2006
    \textit{ApJ} \textbf{641}, 1051.
  \bibitem[Elmhamdi et al. (2003)]{Elm03}
    \textsc{Elmhamdi, A., et al.} 2003
    \textit{MNRAS} \textbf{338}, 939.

 \bibitem[Elmhamdi et al. (2006)]{Elm06}
    \textsc{Elmhamdi, A., Danziger, I. J., Branch, D., 
    Leibundgut, B., Baron, E.
   \& Kirshner, R. P. } 2006
    \textit{A\&A} \textbf{450}, 305.

\bibitem[Fesen,  Kirshner \& Becker]{FKB88}
\textsc{Fesen, R. A., Kirshner, R. P. \& Becker, R. H.} 1988 
In
\textit{Supernova Remnants and the Interstellar Medium}
(ed. R. S. Roger, \& T. L. Landecker) p.\ 55.  Cambridge.  
    
  \bibitem[Gaensler et al. (1997)]{Gae97}
    \textsc{Gaensler, B. M., Manchester, R. N., Staveley-Smith, L., 
Tzioumis, A. K., Reynolds, J. E. \& Kesteven, M. J.} 1997
    \textit{ApJ} \textbf{479}, 845.

  \bibitem[Grasberg et al. (1971)]{GIN71}
    \textsc{Grasberg, E. K., Imshenik, V. S. \& Nadyozhin, D. K.} 1971
    \textit{Ap. Sp. Sci.} \textbf{10}, 28.

  \bibitem[Hendry et al. (2006)]{Hen06}
    \textsc{Hendry, M. A., et al.} 2006
    \textit{MNRAS} \textbf{369}, 1303.
 \bibitem[Laming \& Hwang (2003)]{LH03}
    \textsc{Laming, J. M. \& Hwang, U.} 2003
    \textit{ApJ} \textbf{597}, 347.

  \bibitem[Hwang et al. (2001)]{Hwa01}
    \textsc{Hwang, U., Petre, R., Holt, S. S. \& Szymkowiak, A. E. } 2001
    \textit{ApJ} \textbf{560}, 742.
  \bibitem[Matheson \& Safi-Harb (2005)]{MS05}
    \textsc{Matheson, H. \& Safi-Harb, S.} 2005
    \textit{Adv. Sp. Res.} \textbf{35}, 1099.
    
  \bibitem[Maund et al. (2005)]{Mau05}
    \textsc{Maund, J. R., Smartt, S. J., Kudritzki, R. P., Podsiadlowski, 
    P. \& Gilmore, G. F.} 2005
    \textit{Nature} \textbf{427}, 129.

  \bibitem[McCray (2005)]{McC05}
    \textsc{McCray, R. A.} 2005
    In \textit{Cosmic Explosions, On the 10th Anniversary of SN1993J} 
    (ed.\ J. M. Marcaide \& K. W. Weiler) p. 77. Springer.

  \bibitem[Nomoto et al. (1982)]{Nom82}
    \textsc{Nomoto, K., Sugimoto, D., Sparks, W. M., Fesen, R. A., 
    Gull, T. R. \& Miyaji, S.} 1982
    \textit{Nature} \textbf{299}, 803.

  \bibitem[Nomoto et al. (1996)]{Nom96}
    \textsc{Nomoto, K.,  Iwamoto, K., Suzuki, T., Pols, O. R., 
    Yamaoka, H., Hashimoto, M., Hoflich, P. \& van den Heuvel, E. P. J.} 1996
    In \textit{Compact Stars in Binaries} 
    (ed.\ J. van Paradijs, E. P. J. van den Heuvel \& E. Kuulkers) 
    p. 119. Kluwer.
    
  \bibitem[Podsiadlowski (1992)]{Pod92}
    \textsc{Podsiadlowski, E. M.} 1992
    \textit{PASP} \textbf{104}, 717.

  \bibitem[Ryder et al. (2004)]{Ryd04}
    \textsc{Ryder, S. D., Sadler, E. M., Subrahmanyan, R., Weiler, K. W., 
   Panagia, N. \& Stockdale, C.} 2004
    \textit{MNRAS} \textbf{349}, 1093.
  \bibitem[Schlegel (1990)]{Sch90}
    \textsc{Schlegel, E. M.} 1990
    \textit{MNRAS} \textbf{244}, 269.

  \bibitem[Serafimovich (2005)]{Ser05}
    \textsc{Serafimovich, N. I., Lundqvist, P., Shibanov, Yu. A. \& 
    Sollerman, J.} 2005
    \textit{Adv. Sp. Res.} \textbf{35}, 1106.
    
  \bibitem[Seward, Gorenstein \& Smith (2006)]{SGS06}
    \textsc{Seward, F. D., Gorenstein, P. \& Smith, R. K.} 2006
    \textit{ApJ} \textbf{636}, 873.

  \bibitem[Soderberg et al. (2006)]{Sod06}
    \textsc{Soderberg, A. M., Chevalier, R. A., Kulkarni, S. R. \& Frail, D. A.} 2006
    \textit{ApJ} in press (astro-ph/0512413).
  \bibitem[Sugerman et al. (2005)]{Sug05}
    \textsc{Sugerman, B. E. K., Crotts, A. P. S., Kunkel, W. E.,
Heathcote, S. R. \& Lawrence, S. S.} 2005
    \textit{ApJS} \textbf{159}, 60.

  \bibitem[Turtle et al. (1987)]{Tur87}
    \textsc{Turtle, A. J., Campbell-Wilson, D., Bunton, J. D., 
    Jauncey, D. L. \& Kesteven, M. J.} 1987
    \textit{Nature} \textbf{327}, 38.

  \bibitem[Weiler et al. (2005)]{Wei05}
    \textsc{Weiler, K.~W., Dyk, 
S.~D.~V., Sramek, R.~A., Panagia, N., 
Stockdale, C.~J. \& Montes, M.~J.} 2005
    In \textit{1604-2004: Supernovae as Cosmological Lighthouses} 
    (ed.\ M. Turatto, S. Benetti, L. Zampieri \& W. Shea.) p. 290. ASP.
  \bibitem[Woosley et al. (1994)]{Woo94}
    \textsc{Woosley, S. E., Eastman, R. G., Weaver, T. A. \& Pinto, P. A.} 1994
    \textit{ApJ} \textbf{429}, 300.
  \bibitem[Woosley et al. (1999)]{WES99}
    \textsc{Woosley, S. E., Eastman, R. G. \& Schmidt, B. P.} 1999
    \textit{ApJ} \textbf{516}, 788.


 
\end{thebibliography}
\end{document}